A Multipoint Method for Approximation of Wavefunctions at a Mesoscopic Scale

M. George
Dept. of Physics
Southwestern College
Chula Vista, CA  91910
mgeorge@swccd.edu
12/4/2014

Abstract

We discuss the concept of a mesoscopic wavefunction, first in a general context, as the concept of wavefunction has evolved, and then in a more specific context of modeling.  The paper concentrates on a simple, abstract one-dimensional situation.  In this context, there are three problems to be considered.  In the first problem, we consider the construction of a wavefunction as a problem of interpolation, with information content in a multipoint perspective at finitely many discrete points and complete uncertainty elsewhere.  The wavefunction is conceived abstractly as our unified subjective picture of information content.  Each point of information is essentially free and independent of all others.  This is a wavefunction for a classical system at the mesoscopic threshold.  In the second problem, we consider how, using the concept of scaling and renormalization, the classical system can be brought  to represent a mesoscopic level of integrated information, with points still treated as free, but now with the need to consider each point as an extended region, with possible boundary overlaps.  We then, in the section on the third problem, consider modeling this renormalized system as an Ising-like system of interacting spins.  This is the final picture we develop for a mesoscopic wavefunction.  This can be viewed from the perspective of the evolution of the concept of wavefunction at the microscopic level, and we briefly discuss the new point of view being developed here.  Finally, we present a discussion concerning the bearing of this on Gibbs phenomenon.

**Introduction**

This paper concerns a number of problems related to mesoscopic systems.  Conceptually, the idea of mesoscopic systems [1] is that of a physical domain one starts to encounter as particle size and particle number gets smaller, as with nanoparticles [2] or quantum dots.  As size is reduced, from a macroscopic scale, quantum effects become significant, surface-to-volume ratio becomes large, and active contact of the system with its environment results in novel effects.  In mesoscopic magnetic systems [2], for example, the Curie temperature is observed to be size-dependent.  A quantum dot [3] is typically a semiconductor material of the size of about 100 nm.  These systems can be thought of roughly as artificial atoms consisting of about 1 million actual atoms, with a corresponding number of electrons.  They are extended coupled quantum systems, that can be understood, in part, in terms of the theory of quantum chaos.  They manifest the quantum Hall effect, and can be studied in the context of non-equilibrium statistical mechanics as time-varying quantum systems.

The mesoscopic systems, while extraordinary and fascinating in the purely physical realm that miniaturization is making available to us, can be broadened to include biological systems [4]. For example [4], DNA can be thought of as a mesoscopic system. Then, we can consider modeling local interactions between proteins and DNA to study the general dynamics of the protein in terms of the conformation of DNA. This type of simulation encounters numerous complicating quantum and classical details and must necessarily neglect many of these, e.g. focusing only on base pair sequences [4]. One can then roughly study the free energy landscape for protein-DNA interactions. A search for the most populous microstate can then identify promoter sites, given specific characteristics of a type of organism, as a bacterium [4], from which data is drawn for the simulation.

An important area of quantum chemistry that is starting to be applied to DNA [35] is the *ab initio* computation of electronic structure. In our paper, we are interested in the construction of a mesoscopic wavefunction, but there is an elaborate combination of theory, computation and experiment that chemists have brought to bear on this problem for molecular systems of small to intermediate size.

Our context for the work presented in this paper is empirical, gradually extending into the mesoscopic domain from a classical or Newtonian framework of physical information. It is not fundamental as in quantum mechanics. Furthermore, we do not address electronic structure. However, the whole concept of a wavefunction is arrived at from quantum mechanics, and it is necessary to discuss the quantum mechanical studies of molecular structure [36], as it is a very significant area of study which is gradually impinging from the microscopic domain on mesoscopic structures, and gives meaning to our concept of mesoscopic wavefunction, as a subjective, unified picture of both information and uncertainty about a physical system.

This microscopic viewpoint is an important context to review, and it is an extremely large and important area of chemistry. Below, we only touch on the gradual evolution of this field. Although well-known, it is important to bear in mind that there are four general branches into which this work has evolved [36]: The semiempirical molecular quantum mechanical methods; ab initio calculations; the density-functional method; and the molecular-mechanics method. The density functional method calculates molecular probability density, and not a wavefunction. The molecular-mechanics method is not a quantum-mechanical method, and does not use the concept of wavefunction. Thus, our mesoscopic wavefunction, which is not quantum-mechanical, represents quite a break from all of these branches, and is, conceptually, a very novel point of view. Although we only consider this theoretically, and very abstractly in one-dimension only, computationally it would be unique as well. Therefore, our point of view represents a different perspective from the four conventional viewpoints we have mentioned. This is not surprising, as the concept of a mesoscopic wavefunction defining structure can borrow characteristics from both microscopic and macroscopic characterizations. The idea even that the concept of wavefunction might be useful, detached from fundamental theory, and as a unified picture of information and uncertainty, abstracts the notion of wavefunction in terms of its application, rather than fundamental meaning.

We now trace the understanding and development, through semiempirical, *ab initio* and density functional computation, of the concept of wavefunction, as it has evolved toward the mesoscopic domain. Then, we consider some of the important models being used, as we use a simple, Ising-like model, which contrasts, but bears similarity with some of the electronic structure models for microscopic systems. As well, the use of models in condensed matter physics focuses on lattice computations, and these can be considered to be direct progenitors to our approach.

Progress has appeared over a period of many years, starting with seminal work at the birth of modern, non-relativistic, quantum mechanics. *We* have to put some perspective on the immense computational progress [44] towards developing mesoscopic methods in chemical physics. This work sets the background for the microscopic theory that is gradually encroaching on very precise mesoscopic computations. Our work, although of a different nature and more mathematically abstract, must be seen against this enormous effort that is being made. Because such tremendous progress has been made, much of it very rapid in recent years, it is a necessary perspective and not a digression. So much work, and so diverse agendas have been followed, that it is difficult to select just a few that bear on our paper.

For the wavefunction, one can start with a simple many-body Hamiltonian, that contains both electrons and nuclei, and uses electrostatic interactions only [36]. The formalism [35] here is elementary and well-known, but we will briefly review it [35, 37] because it is important to be aware of the various approximations that go into constructing the straightforward microscopic theory that is used to build toward a mesoscopic regime. In fact, it is satisfactory to use, to some extent, just the formalism for the electronic structure of a diatomic molecule. One then uses the Born-Oppenheimer approximation to obtain a purely electronic Hamiltonian, with fixed nuclei. Then, the Hartree approximation can be applied for the electron-electron interactions, replacing them by a single-electron interaction in a self-consistent field, from an electronic sea from the other electrons. The wavefunction at this point is a Hartree representation that is not antisymmetric with respect to exchange. Our mesoscopic wavefunction, because it is empirical, makes no effort to appeal to electronic structure or exchange either. Leaving out this microscopic symmetry seems appropriate, until we have made further progress in developing the ideas in this paper. In addition, out wavefunction is not associated with a Hamiltonian, but represents information content for a mesoscopic system. This is very important to emphasize that we view the wavefunction structurally and abstractly as a unified representation of both information and uncertainty. Again, in this abstract context, it seems appropriate to avoid the concept of Hamiltonian altogether. Nevertheless, in an information context, the notion of wavefunction for a mesoscopic system is not without meaning.

The next approximation made is that of Hartree-Fock. This is based on the Hartree representation, but uses an antisymmetric wavefunction with respect to exchange. A total energy is computed that leads to Coulomb integrals and exchange integrals. Energy minimization leads to the Hartree-Fock differential equations. These inter-related equations must be solved on the basis of self-consistency. Correlation effects among the electrons are not accounted for, and there are various methods that do this. Some of the methods [43, 44] to account for correlations are the Møller-Plesset perturbation theory [38, 42]; the configuration interaction method [39]; coupled cluster method [40]; and density functional theory [41].

It is important to mention this, as our statistical mechanical approach, using information, constructs an Ising-like model. The model preserves some of the structure from the quantum mechanical models that incorporate exchange. This is a concept related to our concept of mesoscopic wavefunction, despite the fact that in an information picture as we will discuss, there is no actual Hamiltonian.

Density functional theory has, as its fundamental object electron density. Due to its modern importance, we will summarize density functional theory, for which the fundamental object is electron density, rather than a wavefunction. This method has, as a progrenitor, the Thomas-Fermi theory for the atom [45, 47], which, like the Hartree method, arose very early in the development of quantum mechanics , and did not account for correlation and exchange. It is worthwhile noting how soon after the publication of Schrödinger's seminal work [46] this occurred. It is also noteworthy for breaking new ground theoretically, despite its lack of predictive success.

Very important progress in using electron density resulted from the basic density functional theory developed by Hohenberg and Kohn [48]. They showed that it ought to be the case that the electron density, a mere statistical quantity, determines ground state properties: total system energy, kinetic energy, potential energy and wavefunction. The ground state electron density determines, through various functional relationships, expectation values for any observable.

A critical aspect of the density functional approach arises from one of the Hohenberg and Kohn results, namely that given the external potential, trial densities and wavefunctions can be used to place an upper bound on the ground state energy [48]. Kohn and Sham [49-53] introduced some important ancillary ideas. In particular, as the kinetic energy is the most significant contributor to the ground state energy, accuracy in computation from density is extremely important, and Kohn and Sham discuss the use of orbitals for this calculation, which would provide potentially high accuracy.

The principal idea proposed by Kohn and Sham [53] has proven to be very important. This is that the kinetic energy for the many-particle interacting system can be computed from an equivalent energy for a non-interacting system so that individual orbitals can be used in the computation. This statistical but accurate approach is based on the spin orbitals with their occupation numbers. The idea extends to the computation of energy, for which the kinetic energy for the non-interacting system is one contributor. Another term has the form of a classical Coulomb energy for electron-electron interactions. Further, there is a term to incorporate exchange and correlation, with respect to the many-particle interacting system. A final term accounts for the attraction to the fixed nuclei.

The cost for replacing the interacting system by an equivalent non-interacting system for computing kinetic energy is a system of non-linear equations called the Kohn-Sham equations. The effective potential used in this system has a partly intuitive structure in that it has a Hartree-like term. The critical term in the effective potential, however, is a term to account for correlation and exchange. In computing time, the cost is quite severe for such a nonlinear system, as it is solved self-consistently, via iteration from an initial guess for the density.

As one might imagine, incorporating correlation and exchange in the nonlinear Kohn-Sham equations is a serious challenge for computation. Nevertheless, the density functional theory has provided one of the principal avenues for the elucidation of molecular structure. There are numerous algorithmic approaches which attempt to address this. One is the local density approximation [51]. In this approximation, exchange is handled rather simply as associated with a homogeneous electron gas. Addressing correlations can require a reliance on perturbation theory or a Monte Carlo technique. With such an approach, one has hopes that the self-consistent solution to the nonlinear system will converge accurately for slowly-varying electron densities. However, the local density approximation fails even for semiconductors and insulators.

Such unusually difficult mathematical environments are not that uncommon even when dealing with fairly simple nonlinear systems [55]. Nonlinearity can pose severe problems for computation, for example in cancellations that occur in the arithmetic processing of terms. However, this situation has been addressed with the local density approximation by explicitly accounting of electron density gradients [54, 56-59]. This method developed into the generalized gradient approximation, and it proved to be an extremely significant advance [60-62]. The local density approximation led to the most popular applications of density functional theory during the '70s and '80s, while by the late '80s, the generalized gradient approximation was leading to results satisfactory for chemical calculations [64]. Later, significant progress on this latter method [65], led to this becoming the most popular approximation used in chemistry today. Most successful algorithms apply either to molecular systems [66] or extended bulk systems [67, 71]. Extension to bulk systems is based on plane waves with periodic boundary conditions [68]. Current algorithms [69, 70] can address either molecular systems or extended bulk systems.

Very recently [63, 72], there has been some interest in renormalization, with density functional theory. Renormalization is one of the main methods we use in our present article, and although density functional theory does not yet reach into the mesoscopic domain, one can see, in Ref. 63, that much progress is being made.

The mathematical models utilizing density formulations and wavefunctions need to be discussed to some extent. In this context, although there are a huge number of numerical algorithms applied to these models, we can only discuss a few.

A basic, very simple model for interacting particles is the Hubbard model [77]. The Hamiltonian has a tight-binding part, with nearest-neighbor hopping, and a potential energy based on density-density Coulomb repulsion. The model can be used to study conduction-insulating transitions [78]. A few of the numerical methods applied to the model are: determinantal Quanum Monte Carlo (dQMC) [79]; Quantum Cluster Approximation (QCA) [80]; and Density Matrix Renormalization Group (DMRG) [74, 81, 82].

Another and related model is the t-J model [73]. It has been useful for numerous studies. The second quantization form of the Hamiltonian has a particularly simple structure. The t-part of the Hamiltonian is the tight-binding part and has a hopping part over nearest-neighbors. The J-part has a Heisenberg, spin-spin interaction term over nearest-neighbor sites. It also has a potential energy over nearest-neighbors based on density. Physically, it is of interest as a strongly-correlated system. A variety of methods have been used to study it, including: iPEPS [73]; and DMRG [74].

Returning to the context of biological systems, the notion of mesoscopic system can even be broadened beyond the level of macromolecules like DNA [5]. In this broader view, a biological system, modeled perhaps as a graphical network, can be seen as generating macroscopic simplicity from microscopic complexity. The suggestion is made [5] to focus on biological regulation as resulting in an overall simplicity in the larger system, somewhat akin to ferromagnetism. There is a multilevel character to modeling biological systems as networks. This focuses not only on the quantum physics of a system, which is the microscopic level of a graphical model, but also more classical, ordered behavior at a macroscopic level. The entire intermediate level, between these two extremes, then, is not simply a small system, but one in which, as we scale up from detailed microscopic complexity, addressable at the level of quantum mechanics, and uncertainties, there is an evolution into classical, and Newtonian behavior.

More generally, the elucidation of mesoscopic physics can be seen as one of measurement, where conflation [6] from numerous physical sources introduces what Ref. 6 refers to as the inverse problem. In this view, experience serves to supply ways of circumventing the inverse problem [6]. Circumventing the inverse problem is what ultimately makes the idea of an empirical, rather than fundamental, approach to wavefunction meaningful.

What constitutes an "intermediate" system, a mesoscopic system? Our concept of mesoscopic systems is broad. The virtue of a broader definition is that it makes one aware of the great potential for new applications. Thus, the multipoint approach we discuss in this paper is very abstract and uses no theoretical frameworks from either quantum mechanics or classical physics. However, our multipoint method for approximation does derive from the procedure of renormalization [7] in statistical mechanics. After we finish discussing these general ideas about mesoscopic systems, we will review the method of approximation discussed in this paper, which is specialized for mesoscopic systems. Following this, we will give an overview of the three problems we address concerning this approximation method, and then discuss Gibbs phenomenon [8]. Gibbs was one of the founders of the science of statistical mechanics, and in Gibbs phenomenon, the familiar overshoot at a discontinuity, some have thought that Gibbs was anticipating the intermediate region between quantum and classical mechanics [8]. This phenomenon is necessarily an artifact of Fourier analysis in certain situations of sharp boundary behavior, but mesoscopic systems bring into focus a somewhat alternative picture. An approach using ideas about the Ising model [9] from statistical mechanics can potentially circumvent Gibbs phenomenon in the case of mesoscopic systems.

We mean by the term mesoscopic system, certain characteristics about the phenomenology of such a system pertaining to measurement. This means that we are not specifically concerned with the intermediate regime between quantum mechanics and classical mechanics. It is the mode of measurement. This will pertain to a broader class of phenomena.

With respect to measurement, we can think in local terms, in the context of space and time, or, we can consider local measurements in the context of wavelength and frequency. These are the two extremes that we wish to juxtapose to understand what the term mesoscopic refers to. For local measurements, in space and time, for example, it is meaningful to speak of position, velocity and acceleration of points.

On the other hand, and at the other extreme of measurement, it is meaningful to refer to spectra, with, for example, interference, amplitude and phase.

Because Newtonian mechanics expresses the first extreme in measurement, and quantum mechanics is a paradigmatic example of the other extreme, it is common to characterize mesoscopic in terms of these theories. However, we wish to focus on measurement and the result of measurement, without reference to a specific theory. Then, we can see the term mesoscopic as referring to a system in which there is both a significant amount of local space and time-related measurements, as well as significant information related to wavelength and frequency.

It is not the size of a system that is so important in this picture, or whether or not Planck's constant must be taken into account, but just purely measurement and the result of measurement to which we refer in defining what we mean by a mesoscopic system. This is why we refer to our approach as empirical, rather than fundamental. On the other hand, a macroscopic system is one in which measurement relates exclusively, or almost so, to local quantities in space and time, and a microscopic system is one in which measurement is exclusive to local quantities in wavelength and frequency.

Thus, the methods we discuss in this paper are oriented toward this empirical perspective, and intended to apply to mesoscopic systems. Such systems are becoming more and more a focus for research attention. As examples, one can cite DNA research [4, 5, 10]. In [10], smooth dissipative particle dynamics was used to study tethered DNA in shear flow. In this case a particle of fluid is related to thermodynamic fluctuations in the fluid. The measurements provide "a priori" characterizations for simulations. Computational requirements for a fully atomistic study are prohibitive. A mesoscopic simulation connecting microscopic to macroscopic scales is a good alternative. With respect to DNA, there is local information carried by atomic structure of introns and exons, but there is a larger scale perspective, related to the scale of genes and chromosomes, for example, in which the spread of mutations and mutation rates, akin to wavelength and frequency, become important concerns.

Recent research in ecosystems [11] can address ecologies in terms of mesoscopic systems. In [11], a non-equilibrium model for ecosystems is addressed based on thermodynamics. In this context, stability of dynamical evolution is studied. In the context of the complex dynamics, arising from numerous interactions, usually expressed in terms of systems of nonlinear differential equations, macroscopic properties of ecosystems have been the focus. The study of local perturbations, using Lyapunov functions, is not sufficient to consider global stability in a number of typical predator-prey models. Construction of a potential energy landscape to study this problem in [11] requires knowledge at a mesoscopic level of stochastic variations using the Fokker-Planck formalism [12, 13] in numerical simulations to study diffusion. The microscopic level of individual members of habitats is inaccessible.

Some research interests in neuroscience [14] have also focused on mesoscopic systems. Using graphical networks, in [14], the authors are able to study diffusion currents in the thermodynamics of non-equilibrium networks. The result is a fluctuation theorem [13] for the current, which they apply to a biophysical model of ion channels in a membrane and electron transport in mesoscopic circuits made of tunnel junctions. In [14], some account must be made of the discreteness of matter from the scale of micrometers down to nanometers. At the mesoscopic scale, one needs to consider a stochastic description, which in [14] takes the form of Markov processes.

Stochastic modeling in thermodynamics, and scaling and power-law behavior are two approaches used in the study of non-equilibrium statistical mechanics of mesoscopic systems [7, 13, 17]. Our attention in this paper is focused on methods associated with scaling and the renormalization group theory from critical phenomena. These methods arise from quantum field theory [19], but they are quite amenable for the situation of mesoscopic systems.

Stochastic modeling in epidemiology has considered the spread of an infectious disease as a mesoscopic system [15]. Fluctuations at the individual or microscopic level in [15] were considered to be inaccessible, but at the intermediate level in the structure of the epidemic, stochastic approaches using the master equation were considered.

The examples are manifold, and not just confined to biosystems. As an application of mesoscopic systems, one can think, for example, of the study of traffic flow [16]. This, indeed, is a very good example to understand what we mean. It is often not feasible to study a traffic system at the level of individual vehicles. An individual vehicle obviously has associated with it local space and time measurement, whereas we understand, in the large scale, that referring to wavelength and frequency measurements, in the context of the measurement of traffic flow is very meaningful. As well, when we shift to the wavelength and frequency measurements, statistical measurements become more important.

Thus, the definition of mesoscopic system need not be strictly size dependent, but can focus on empirical properties of the individual vs. the whole, as an intermediate regime. This empirical perspective does not omit any of the conventional small mesoscopic systems. Our emphasis is on scaling, a recursive procedure. This is important for the comprehension and development of a theory of mesoscopic systems, because we can alter the scale, and therefore the wavelength of our measurements to make local space and time measurements at different scales. In this context it is important to consider the physical limitations of scaling [7].

In very simple terms, we can comprehend that the "arithmetic" of scaling is not one that is simply understood in the ordinary context of the four basic numerical operations of addition/subtraction and multiplication/division. Let us suppose we have two real quantities, x and y, where we can think of x as some time of observation, and y as a related measurement, say of position. We may be considering linear models for this system, and so establish our measurement system such that, locally, $y = \lambda x$, is a good fit to data. If we change scale, say $y = sy'$, where $y'$ is linearly related to the old scale through the factor s, then we obtain $y' = (1/s)\lambda x$, which is a simple multiplicative relation, but the concatenation constitutes, in actuality, also a function composition. This is subtle, but in more complicated examples, is more obvious. Thus, scaling is oriented not to the traditional arithmetic operations, but to an "arithmetic operation", function composition, i.e. a recursive, not recurrence, procedure. Unlike the traditional arithmetic operations, recursive procedures are often not commutative, like addition and multiplication, and also, the "inverse operation" quite often, does not exist. Thus, the arithmetic of scaling results in a shift in viewpoint with respect to measurement. This has been well-explored and developed as the renormalization group theory [7] in the context of the study of critical phenomena in statistical mechanics. We are merely trying to extend this theory a little, in studying the three problems we discuss in what follows.

In a sense, the renormalization group theory, although principally associated with quantum field theory and critical phenomena [19], is a theory of mesoscopic systems, because near a critical point, recursive scaling [17] can be studied over a broad range. In our paper, we distinguish slightly between recurrence and recursion, whereas conventionally both are merely taken as indicative of iteration. Both refer to utilization of self-similarity, associated with arithmetic relationships, such as in the master equation [13], but we wish to apply the term recurrence to self-similarity expressed through addition (as in the simple Euler method of differential equations) or multiplication, while we choose to think of recursive as the type of scaling and renormalization used in renormalization group theory. This is a technical distinction we use in this paper. As a recursive theory, renormalization group theory is a type of theory in which we consider no limitations to scaling, and therefore, macroscopic systems and microscopic systems are dealt with, in this theory, only at the extreme limits as we scale to zero or infinity.

Artificial limits can be imposed, but there are no "natural" cutoffs [13] for the theory. In addition, while this theory was developed on abstract models, it is definitely oriented toward measurement and actual physical systems [18]. In actual physical measurements, such as that of neuronal activity [18], finite-size effects are clearly noticeable.

Once one characterizes a mesoscopic system, the question immediately arises as to how to describe such a system within the context of renormalization. In this context, we have considered just a single dimension, and address this using polynomials and series. In this way, we are exposing some of the bare mathematics involved, in a very simple way, in order to better understand. Local space-time information can then be represented as low-degree Taylor polynomials, and the local wavelength-frequency information can be thought of as represented through infinite series. This allows us to study some very simple problems, which, while not providing a coherent multipoint method for approximation in modeling of mesoscopic systems, allow us to explore a few important mathematical aspects of what is involved. We have chosen these simple problems to develop understanding. It is important to emphasize that they only represent a first step toward an encompassing multipoint method, in analogy to quantum field theory, for modeling mesoscopic systems.

As we have mentioned, we address three different problems in this paper, at gradually expanding level of difficulty. The first problem views the wavefunction as arising out of a unification of Taylor polynomial expansions at specific points. Each Taylor polynomial is a discrete packet of information. Basically, what we want to do here is sew together a finite, possibly large, number of low-degree Taylor polynomials. To a significant extent, this can be mistaken for an interpolation scheme. If we carry out this unification by using just the information from the polynomials, and constructing another (possibly very high degree) polynomial, this finite series cannot develop periodic properties to represent specific wavelengths or combinations of wavelengths. Instead, despite well-behaved regions in the vicinity of the generating points of the Taylor polynomials, it will of course display significant instabilities, relative to representing periodic functions. We develop an approach that involves the representation of uncertainty, and does control instability, but is nevertheless still unsatisfactory, in terms of leading to periodic functions. Although we did not produce a "mesoscopic wavefunction", there are some aspects of interest in the solution we present to this problem. It seems to be a first step in development of a wavefunction, that might eventually be useful in modeling from empirical data.

Before getting into the discussion of the first problem it is worthwhile to paint a fuller picture of what is involved. Often in modeling we think of using exponential functions, or "exponential-like" functions such as Fourier series of sinusoids. However the actual mathematical meaning behind these functions is tied to limits of polynomial sequences, say Taylor polynomials. The convergence of these polynomial sequences obscures the natural instabilities inherent in polynomial sequences, and possible chaotic behavior that can occur outside sharp convergence regions on compact intervals. Incidentally, this harks back to some of the classic work of pure mathematics [20, 21], here being applied in the context of mesoscopic system theory. However, this classic work often focuses on the concept of "summability". What we wish to emphasize, which is quite different, is that outside of very limited domains, sequences of polynomials can behave very chaotically, and represent a loss of information. There is nothing to be summed or any meaning to be derived from these regions. These regions are, simply, domains of uncertainty. We wish to represent this loss by boundary behavior with the regions on which good convergence is being observed. It is this duality, between information representation and uncertainty, which distinguishes our approach to the first problem from a mere interpolation scheme. It is more appropriate, and more tied to the meaning of the concept, to refer to the unified picture as a "wavefunction".

Although we use analytic functions, and this type of analysis encourages that [22], we wish to emphasize that renormalization focuses precisely on non-analytic behavior. This is why we do not discuss, to any extent, analytic continuation. It is not a matter of discounting the conventional theory of analytic functions of a complex variable, merely that one must also consider the possibility of non-analytic continuation. We emphasize, for example, that although we only use analytic functions, $C^\infty$ functions may be of interest in general which may not be analytic.

Periodic functions like sine possess clear scaling behavior, as a result of intrinsically possessing "correlations" at all length scales, as in critical phenomena [19]. Rather surprisingly, this means that the very well-understood and familiar trigonometric identities can be used to devise renormalization transformations. Following the polynomial-like representation of information that we discuss in the section on the first problem, this becomes an extremely key observation, because it shows that a mesoscopic wavefunction can be potentially devised from the representational structure of the local information, by scaling that local information outward. Thus, an alternative to the approach of the first problem is that focused around a second problem. This second problem, while not connecting to non-equilibrium transport theory, does pose, as a goal, a dynamics, structured around the information-rich multipoint approximation devised from local information of Taylor polynomials. The second problem constitutes how to use local information from Taylor polynomials to develop scaling extrapolations at each generating point of a Taylor polynomial, that extend the point information into a whole region. These are not renormalization transformations, but they bear a close resemblance to these, and permit the evolution of relatively stable structures, from local kernels, centered around the generating points of Taylor polynomials. These extended structures ignore detail, have overlaps with adjacent structures, and cannot be expected to detect sharp boundaries.

While this theoretical approach presents a possibility of yielding what one might think of as mesoscopic wavefunctions, a third problem arises concerning the overlap between extrapolations and the detection of sharp boundaries.  This is the analogue of considering phases of sinusoids for interference, which we know is a critical part of quantum mechanical wavefunctions.  There is a more meaningful idea surrounding phase at the mesoscopic level, than at the microscopic level.  The lack of direct physical measurement for a quantum phase is expressed in a rather meaningful ambiguity about the overlap resulting from the multipoint construction discussed in the section on the second problem.  Certainly connecting the renormalization group approach that we adopt to transport theory requires some resolution of this.  The abstract, one-dimensional mathematical environment in which we operate precludes this.  The mesoscopic domain is an intermediate region, and is involved in what might be called the "classical limit" of quantum mechanics.  Our approach to this is very different from conventional discussions of the classical limit, because we are connecting, and taking the first steps toward, a master equation at the mesoscopic scale, rather than what is commonly taken as the classical limit.  This involves a whole developing field in quantum mechanics, where investigators are utilizing the correspondence between imaginary time and temperature [23].  This is a rich subject, and one worth pursuing, but is somewhat of a diversion for us in this paper on multipoint approximation.  Therefore, although the classical limit, as a connection with transport theory, is of great interest, we focus on the mathematical problem of overlaps and boundaries, that arise in discussing the second problem.

This is taken up as a third theoretical problem.  With this, we essentially describe a multipoint method for approximation that is the goal of this paper.  The proposal is to develop Ising-like models [24, 25] of interactions between neighboring "spins", where the "spin" amounts to the Taylor polynomials, i.e. interactions are between Taylor polynomials.  In previous work [25], we considered Padé approximants to model mesoscopic phenomena, and our multipoint approach in this paper corresponds, to some extent to multipoint Padé approximants [26].  This statistical mechanical method might then lead to resolution of detail involved with overlap and sharp boundaries as domain boundaries of the Ising-like model.  While this provides only a theoretical framework, it seems to be one that takes a first step toward description of mesoscopic systems, both in terms of a mesoscopic wavefunction, and through an interaction model.  This theoretical framework proposes a new type of method for approximation of functions that are defined from empirical data on mesoscopic systems.

One important idea that comes out of focusing on mesoscopic systems is that it might be feasible to avoid Gibbs phenomenon [27] in considering sharp boundaries.  This is discussed in the final section of this paper prior to concluding remarks.  Boundaries appear as domain boundaries in the Ising-like model, instead of resulting from Fourier analysis of waves.  Enormous progress has been made, to circumvent Gibbs phenomenon in Fourier analysis, e.g. in the analysis of MRI data [27].  Nevertheless, our approach is very different and may be of some interest.

The outline of the paper is as follows:  In the next three sections, we discuss, respectively, the three problems, developing the framework for the multipoint method of approximation in the section on Problem 3.  There is a discussion section following this, which considers the nature of data defining functions for a mesoscopic system.  The section following this considers Gibbs phenomenon, and then there is a concluding section.

**Problem 1**

The first problem we consider is to find a function which, when expanded in a Taylor polynomial at each of several points, will reproduce the Taylor polynomial to the specified degree at each point. And the overall goal of this exercise is to represent the information contained in the separate Taylor polynomials in this unified way, while also representing the lack of information and uncertainty at other points. We could consider this as an interpolation problem for splines and Bezier curves. But this would not be satisfactory. We must also consider the representation of the uncertainty or lack of information between generating points of the separate Taylor polynomials.

The problem we are addressing is not the smoothing of data or the interpolation of function information or fitting. Our problem concerns the representation of information, i.e. we want not only to indicate presence of good approximating information, but the absence of information or uncertainties. Therefore, splines [28, 29] are not a device that is appropriate for what we are considering here. Rather, we are considering the construction of a mesoscopic wavefunction [33, 34], i.e. an object that is carrying the physical information contained in the Taylor polynomials, as well as indicating uncertainty elsewhere. Similarly, Bezier curves [30] would not be appropriate.

Our solution to this problem involves, at least at an initial level, an expectation of high degree polynomials to present a unified picture. Thus, while in the vicinity of a Taylor polynomial, the unifying function will approximate well, as the "edge" of the approximation is approached, the high degree will result in a "polynomial wiggle", and lack of information. Usually, one must say, almost always, perhaps, polynomial wiggle is undesirable. One could cite numerous examples of attempts to eliminate polynomial wiggle, but here is one: [31]. In our case, the polynomial wiggle is a desirable feature because it indicates uncertainty.

One can consider what this means for a polynomial representation in the region between adjacent generating points of Taylor polynomials. If the points are close (we will not try to define "close" although it means, in this context, loosely, moderately separated on a microscopic scale, but very close on a macroscopic scale, i.e. at the microscopic edge of mesoscopic phenomena) then the two Taylor polynomials incorporate well both generating points within the region in which they are good approximations. In such a case, we would notice no region of uncertainty.

It is important to understand, before we actually construct a unifying polynomial, why this is, even though it is an elementary and straightforward fact. The Taylor polynomials provide information at relatively low degree, at least in relation to the unifying polynomial. Therefore, the rapid variations inherent in high degree (somewhat complex) polynomials that can lead to significant boundary instability, are not noticeable. This is a result of an elementary property of polynomials: High degree polynomials display boundary "instabilities". This is actually noticeable even for fairly low degree polynomials when we perform fits and interpolation (see [31]), and it constitutes what is observed as the notorious polynomial wiggle. This is contained in the observation that while $y = x$ displays "smooth" behavior throughout its domain, $y = x^5$ is rather flat, contributing values close to zero, in a neighborhood of zero, while its behavior starts to change dramatically around x = 1, and for x > 1 there is a steep increase in this function, i.e. the high degree part of the unifying polynomial is "asleep" in the vicinity of the region about the generating point where the low degree Taylor polynomial is a good approximation. As we move away from this region, the high degree parts of the unifying polynomial start to "wake up", actively introducing instability. Of course, it must be born in mind that we are merely speaking loosely, and not trying to precisely define what we mean by low degree, etc.

What is the situation if the generating points of two adjacent Taylor polynomials are very far apart on a microscopic scale but also moderately separated on a macroscopic scale? This is the situation at the macroscopic edge or limit of the mesoscopic regime. Many of the very high degree terms in the unifying polynomial have "awakened", and they are outside their region of correlation, when we consider a point in the intermediate region between the two generating points. Furthermore, these high degree terms are only well-correlated toward the "boundaries" (if at all) where they start to become active in the regions where the Taylor polynomials form good approximations. We assume we are outside those regions for both polynomials. There is a great deal of instability, and a significant amount of polynomial wiggle, driven by the rapidly increasing or decreasing behavior of these terms.

The reason for the observed polynomial wiggle is elementary and simple, but, nevertheless, worth stating explicitly. Although the high order terms may (or, actually, may not, since there is an enormous amount of flexibility and ambiguity in the construction of a unifying polynomial) be highly correlated with Taylor polynomials generated at numerous points away from the two we are considering, they behave, in this intermediate region we are considering, in an essentially uncorrelated and random way. (We assume also that the type of scaling we discuss later cannot be applied in this region: we are at the edge of mesoscopic behavior. Discussions of scaling extend back many years [32] in the field of particle physics. At that time, the cut-off regions were generally taken as sharp, and the possibility of mesoscopic behavior was not considered.)

The result of a lack of correlation, as well as being in an intermediate region for the unifying polynomial, leads to the observed oscillations of polynomial wiggle. For example, consider $x^3/12! - x^4/16!$ (we do not have to consider terribly high degrees to see the result of lack of correlation). One must imagine that the signs of the coefficients here are random, as well as the magnitudes, 12! And 16!. The steep rise in $x^3/12!$ Just starts at about $\sqrt[3]{12^{12}/e^{12}}$, from roughly applying Stirling's formula, as we move away from x = 0. This quantity is about: 380. On the other hand, the steep decrease in the quartic term just begins to occur about at: 1200. What this means is that the cubic term with its increasing behavior, might dominate beyond about x = 380, and then, there would be a gradual turnaround somewhere out about x = 1200. This gives a rough idea how these oscillations or polynomial wiggles occur. And this is despite the fact that these two terms, elsewhere, might contribute highly correlated contributions to some Taylor polynomial.

The instabilities, including polynomial wiggle, for the unifying polynomials in the intermediate regions can be astonishing. This is due to the very rapid increase or decrease in high order terms, away from generating points of the Taylor polynomials. From the viewpoint of the unifying polynomial as representing a mesoscopic wavefunction, we are witnessing, in such behavior, an indication of lack of correlation at a certain point, due to lack of information, despite the fact that elsewhere, the high order terms may supply contributions which are highly correlated and not random at all.

This explores that what we are discussing here is representation of information. It would be a mistake to suppose we are attempting to interpolate, make some kind of smoothing fit, or make the resulting function, i.e. the unifying polynomial, "nice". It's bad behavior importantly signifies loss of information.

The problem of devising a unifying function, is certainly the easiest of the problems we consider, and we will use, to start, polynomials, as we have discussed above, of relatively high degree. The reason for this is that, in general, the behavior of a polynomial at one point can be very independent of its behavior at another point, particularly if the points are not particularly close to each other. This means that the resulting polynomial "unifying" the Taylor polynomials at several points, can have degree much higher than any of the individual Taylor polynomials.

Now, let's get down to presenting the details of a solution. We need only provide a specific construction to display the method. Our initial approach is a simple extension of Lagrange interpolation. The solution can be presented adequately in a specific illustrative case. We will present the example of this in such a way that generalization is clear.

We will consider Taylor polynomials coming from the rational function

$$1 \qquad y = 1 + (1/2)x^2 + (1/6)x^3 - (1/2)x^4 + 1/(x - 4) - 5/(3x^2 + 22)$$

We wish to consider the Taylor polynomials of degree 3, at x = -1, 0, 1, 2, and 3. Degree three is small enough to illustrate simply, and yet large enough to see the relevant problems involved.

We have chosen this function, a rational function, for the simulation, because its structure is easy to visualize, and yet it is not quite in the category of trivial cases, viz. polynomials. Degree three Taylor polynomials will display a small region of good approximation about each of their generating x values, and yet the values x = -1, 0, etc. are sufficiently separated that the polynomials will be relatively independent, and, using just a unifying polynomial that agrees with the Taylor polynomials at each of these points, we will see a substantial amount of instability in the polynomial between the points and outside the range of the x values from -1 to 3. Certainly, this is the characteristic "polynomial wiggle".

In using (1) to develop the Taylor polynomials as a simulation, and then compare with the unified information function, we are faced with the following characteristics. For Eqn. 1, there is a background polynomial of degree 4, a simple pole at x = 4, and a complex pair of simple poles at x = $\pm\sqrt{22/3}$i. The simple pole, at 4, placed so close to the x values where we are interested in Taylor series, of course, gives us a little challenge. We cannot expect for our unifying function, anything comparable to Padè approximants [26] for locating the singularity positions.

At degree 3, for the Taylor polynomials, we have just 4 terms to work with. No obvious way, with such few terms, presents itself, as with Padè approximation, for nailing down singularity positions. Our goal is to record the information of the Taylor polynomials in a unified function, and not supply information about singularities. Thus, although this simulation is very simple, it will supply a wavefunction that appropriately captures point knowledge from the Taylor polynomials, and also uncertainties. Padè approximation is not quite suitable for this because we use it to extrapolate and capture information about singular structure. However, Padè-like rational functions could be used to construct a wavefunction. Our approach, in fact, uses simple Padè-like functions, but we did not use this to create the wavefunction in this case, because we simply want to display the method, and our basis for simulation, Eqn. 1, is not complicated enough for Padè approximation.

The displayed rational function in (1) is expanded about each point x = -1, 0, …. An expansion about x = -1, for example, takes (1) in an expansion about x = -1, starting from:

2 $\qquad y = 1 + (1/2)((x + 1) - 1)^2 + (1/6)((x + 1) - 1)^3 - (1/2)((x + 1) - 1)^4 + 1/((x + 1) - 5) - 5/(3((x + 1) - 1)^2 + 22)$

decomposed into background polynomial and a rational function. An important symmetry to note here, with respect to this decomposition, is that after simplification, the background polynomial is the same background polynomial, with respect to the decomposition at x = 0, i.e. Eqn. 1, and the pole positions have shifted but we obtain the same functions expressing each pole. We are just shifting the position of the origin, so we must have this symmetry, but it is important to point out, because we need to consider Taylor polynomials not only about x = 0, but about other points, including x = -1.

Although we are going to consider the rational function (1) as our example, a function that is a little more interesting might be something like:

3 $\qquad y = (1 + (1/2)x^2 + (1/6)x^3 - (1/2)x^4) \exp(-x^6) + (1/(x - 4)) \exp(-1/(x - 4)^2) - (5/(3x^2 + 22)) \exp(-1/(3x^2 + 22)^2)$

With respect to perturbations, the exponential functions have no effect on the leading orders. However, the exponentials modulate polynomial wiggle and instabilities. This function, in fact, has the kind of structure we wish for a mesoscopic wavefunction representation of (1).

Recall that the Lagrange polynomial interpolation [29] fits neatly into a vector space format, where the elements in the space are just polynomials, and one constructs a very simple basis set associated with the particular points (here x = -1, 0, 1, 2 and 3.) There is little more involved here in constructing a wavefunction, but we are not intending this as an interpolation.

We want a polynomial that takes the value 1 at 0, and equals 0 at x = -1, 1, 2 and 3:

4 $\qquad y = (x + 1)(x - 1)(x - 2)(x - 3)/[(-1)(-2)(-3)]$

Set $x_{-1} = -1$, $x_0 = 0$, $x_1 = 1$, $x_2 = 2$ and $x_3 = 3$, then we can rewrite (4) as

5 $\qquad u_{L0}(x) = (x - x_{-1})(x - x_1)(x - x_2)(x - x_3)/[(x_0 - x_{-1})(x_0 - x_1)(x_0 - x_2)(x_0 - x_3)]$

For the values x = -1, 0, 1, 2 and e, we will have analogous functions to (5) of $u_{L,-1}$, $u_{L,0}$, $u_{L,1}$, $u_{L,2}$ and $u_{L,3}$, and they will be linearly independent. This is the simple vector space structure for Lagrange interpolation.

What is the subspace they generate? It is a certain subspace of polynomials that allow us to find, very efficiently, a polynomial, P, that has specified values at x = -1, 0, 1, 2 and 3:

6 $\qquad P = Au_{L,-1} + Bu_{L,0} + Cu_{L,1} + Du_{L,2} + Eu_{L,3}$

This polynomial has the value A at x = -1, B at x = 0, C at x = 1, D at x = 2 and E at x = 3. Furthermore, it will be, by construction, one of the lower degree polynomials with this property.

Now, having reviewed this, the production of a polynomial that matches Taylor polynomials, that include derivative values, rather than simply function values, is pretty straightforward. For third degree Taylor polynomials, we want polynomials similar to the $u_L$'s. The only difference is that we must specify derivative values as well (through the third derivative). This essentially solves problem 1, as we will see.

The general approach can be seen clearly in this specific example. Let's give an example by constructing a polynomial whose second derivative at x = 1 is 1 and all other values for derivatives (counting function value) up through order 3 at x = -1, 0, 1, 2 and 3 yield 0. The most straightforward way does not work:

7 $\qquad y = (1/2)(x-1)^2(x+1)^4 x^4(x-2)^4(x-3)^4/[(1+1)^4(1+0)^4(1-2)^4(1-3)^4]$

However, it does not take much effort to find a such a polynomial.

The problem here is that y'''(1) does not equal zero. Let us consider a polynomial:

8 $\qquad p_{-1}(x+1) = A(x+1)^4 + B(x+1)^5 + C(x+1)^6 + D(x+1)^7$

We want a "generic" polynomial: Not necessarily a polynomial of lowest degree. We want the polynomial $p_{-1}$ to satisfy:

9 $\qquad p_{-1}(2) = 1$, $p_{-1}'(2) = 0$, $p_{-1}''(2) = 0$ and $p_{-1}'''(2) = 0$

This pushes up the degree of the polynomial a little beyond the minimum, but it supplies a general factor that we can use to write out any of the basis polynomials, the analogue of what happens with Lagrange interpolation, we need in this case. Indeed, the results of the construction are very reminiscent of Lagrange interpolants, except that we are incorporating derivatives. This is the framework of a solution to Problem 1 that we are proposing: the basis for construction of a mesoscopic wavefunction. Note that there is no unique solution (there are infinitely many basis sets that can be constructed, since we are not restricting polynomial degree). The one we propose is simply straightforward and utilitarian.

An explicit solution to system (9), given form (8), is

10     $p_{-1}(x+1) = (35/16)(x + 1)^4 + (-21/8)(x + 1)^5 + (35/32)(x + 1)^6 + (-5/32)(x + 1)^7$

First we will resolve the problem that was not quite solved by (8), namely to find a polynomial such that its second derivative at x = 1 is 1, and all other values, up through third derivatives, at x = -1, 0, 1, 2, and 3 are zero. Based on (10),

11     $p_0(x) = p_{-1}(2x) = (35/16)(2^4)x^4 + (-21/8)(2^5)x^5 + (35/32)(2^6)x^6 + (-5/32)(2^7)x^7$
12     $p_2(x - 2) = p_{-1}(-2(x - 2)) = (35/16)(2^4)(x - 2)^4 + (-21/8)(-2^5)(x - 2)^5 + (35/32)(2^6)(x - 2)^6 + (-5/32)(-2^7)(x - 2)^7$
13     $p_3(x - 3) = p_{-1}(-(x - 3)) = (35/16)(-1)^4(x - 3)^4 + (-21/8)(-1)^5(x - 3)^5 + (35/32)(-1)^6(x - 3)^6 + (-5/32)(-1)^7(x - 3)^7$

Thus, a polynomial that resolves the problem, instead of (7), is

14     $y = (1/2)(x - 1)^2 \, p_{-1}(x + 1) p_0(x) p_2(x - 2) p_3(x - 3)$

This polynomial has a relatively high degree. We were not trying to produce a polynomial of lowest degree, merely give an example. On the other hand, the method we present here is simple, straightforward, and easily generalizable. The point we must emphasize here is that the same polynomial template, $p_{-1}$, is applicable to every case.

To show how to apply this technique in general, let's now find a polynomial such that its first derivative at x = 3 is one, and its other values and derivatives (through the third) are zero at x = -1, 0, 1, 2, and 3. We just need to devise a simple variation on $p_{-1}$ from (10):

15     $q_{-1}(x + 1) = (35/16)(2^4/4^4)(x + 1)^4 + (-21/8)(2^5/4^5)(x + 1)^5 + (35/32)(2^6/4^6)(x + 1)^6 + (-5/32)(2^7/4^7)(x + 1)^7$
16     $q_0(x) = (35/16)(2^4/3^4)x^4 + (-21/8)(2^5/3^5)x^5 + (35/32)(2^6/3^6)x^6 + (-5/32)(2^7/3^7)x^7$
17     $q_1(x - 1) = (35/16)(x - 1)^4 + (-21/8)(x - 1)^5 + (35/32)(x - 1)^6 + (-5/32)(x - 1)^7$
18     $q_2(x - 2) = (35/16)(2^4)(x - 2)^4 + (-21/8)(2^5)(x - 2)^5 + (35/32)(2^6)(x - 2)^6 + (-5/32)(2^6)(x - 2)^7$

Then, the polynomial that satisfies the desired conditions (first derivative at x = 3 is one, etc.) is

19     $y = q_{-1}(x + 1) q_0(x) q_1(x - 1) q_2(x - 2) \, (x - 3)$

This polynomial has a value for a first derivative at x = 3 equal to 1, and function and derivative values (up through third order) equal to zero at x = -1, 0, 1, 2 and 3.

The universality of the structure we are using to find basis functions (analogous to what appears from Lagrange interpolation), makes construction of such functions (which, we emphasize, once again, are not unique) straightforward and efficient. We will use a standard notation, for example $u_{bbb0b}(x)$, to denote elements of the basis set. In the case of $u_{bbb0b}(x)$, this polynomial has the value 1 at x = 2, and the value 0 for all other function values and derivatives (through order 3) at x = -1, 0, 1, 2 and 3. The polynomial $u_{b3bbbb}(x)$ has the value 1 for its third derivative at x = 0, and the value 0 for other function values and derivatives (through order 3) at x = -1, 0, 1, 2 and 3.  The above technique provides one very easy way to construct such functions.  This, in fact, demonstrates how easy it is to construct a mesoscopic wavefunction of the type desired in Problem 1.

This provides an efficient (Lagrange-like function) solution to the problem of finding a polynomial that matches the Taylor polynomials of a function f(x) at x = -1, 0, 1, 2, and 3.  The polynomial is just a linear combination of the basis polynomials, the u's, that we discussed above.  We sum $f(-1)u_{0bbbb}(x)$, $f'(-1)u_{1bbbb}(x)$, $f''(-1)u_{2bbbb}(x)$, etc.  This completes our discussion of problem 1, insofar as the construction of the polynomial is concerned.  We can combine this technique with exponential functions (as we have previously discussed) to obtain a more stable type of function than a polynomial, better representative of the given information, plus what is unknown.  This may be highly desirable, because, while the function we used for this simulation may vary simply over the domain from x = -10 to 10, say, the polynomials can be pretty wild.  In fact, by using such exponential factors, we can construct an adequate multipoint approximation directly from the Taylor polynomials at the various points.  As a remark, we point out that Padè approximation can easily and straightforwardly be generalized as well to this more general type of function.  (Instead of using rational functions, like (x + 1)/(2x – 5), which might be a Padè approximant, we would consider $(u_1(x) + u_0(x))/(2u_1(x) – 5u_0(x))$, where the u's are functions of a more general type as a mesoscopic wavefunction.)

**Problem 2**

Problem 1 addresses the multipoint problem posed in constructing a mesoscopic wavefunction:  One is given separate pieces of information at different points and attempts to unify them, so that the point information is adequately represented, as well as uncertainty between these points.  This involves, usually, fairly high degree polynomials, or somewhat complicated types of functions.  There is, by no means, just one solution to this problem.  We discussed a fairly simple, direct solution to this problem.  This second problem addresses developing a wavefunction that extends outward, based on the point information.  The local information provided at specific points:  How can it be extended?  The point is that we wish to develop an overall picture of the function, but this picture can be supplied potentially by the separate pieces alone.  Problem 2 is to find a way to utilize local information, without unifying it, to supply a global picture of a wavefunction that extrapolates a "likely" picture into the regions that in problem 1, would simply be regarded as uncertain.  This is intuitively possible because each wavefunction constructed from the methods discussed above, not only carry point information, but because a Taylor polynomial is usually accurate in a domain about the generating point, there is a natural framework for extrapolation.

We want to approach this problem by converting Taylor polynomials at various points into information about Fourier components of the function.  The point information, concerning very high degrees in the Taylor polynomials, carry local or microscopic information.  We assume that at low orders, we are obtaining mesoscopic information, i.e. information that can be scaled outward from the generating point of the Taylor polynomial to a certain extent.  A Fourier component can be scaled, and therefore access local information of the Taylor polynomial.  This local information is within the stable region of the polynomial.

In this problem, we concern ourselves only about how a specific component scales. Therefore, we are treating the components independently, and do not attempt to take into account nonlinear behavior between components. In the third problem, which is discussed later, we try to construct a model to account for nonlinearity, or interactions between the regions of point information. This study of scaling is distinctly different than attempting to find an interpolating approximation. We do not attempt to fit the extrapolations: There will be boundary overlaps. Our concept of mesoscopic wavefunction is that it not only carries point information, but preserves uncertainties. This is not a fit to data.

A Fourier component is just a sinusoid and therefore very-well behaved. Furthermore, because it is repetitive, if we know its behavior in the vicinity of local points, where we have information about Taylor polynomials, we may know quite a bit about it. In fact, we will see that special properties of trigonometric functions permit us to derive an enormous amount of information from local behavior.

It may seem that to relate Fourier components to function behavior, we require interpolated or global knowledge. On the other hand, a function's behavior, based on its local behavior, need not necessarily depend on any particular interpolation. As we have discussed, in presenting our resolution to the first problem, the extension from local behavior of Taylor polynomials is by no means uniquely defined, a reflection of uncertainty. Instead, it is important to regard the local behavior as seeds for the growth of a predictive model of global behavior. If this growth can be defined in a natural way from the Taylor polynomial at a point, we can scale up the behavior we see predicted by the Taylor polynomial. This will not generate polynomial behavior as the prediction, but will provide a bootstrap procedure from local data, much more conducive to periodic behavior. In predicting such "spectral" behavior from local conditions, we will produce a resulting picture of somewhat linear-like overlaps of Fourier "extrapolations" from each point, and these will produce, most likely, boundary layers, representing nonlinear transitions from interactions we are not trying to account for, rather than the linear-type extrapolations. Thus, while our resolution to this second problem may seem to circumvent the need for a global interpolant, as we considered in the first problem, we arrive at rather well-behaved extrapolations of a nonpolynomial type (linear and based on Fourier series). This is going to eventually lead to consideration of boundary layers between these well-behaved regions, and take us to the third problem. However, for the purposes of this problem, our goal is to transform local information from Taylor polynomials to extrapolations about Fourier series. We do this by a method of scaling-up the local behavior. Thus, we must examine what we can expect from scaling behavior of simple trigonometric functions, where scaling is simple and well-defined.

Our example is the sine function, which illustrates the general principle. The function $y = \sin(x)$ is an example of a very well-behaved function, throughout its domain. On the other hand, a Taylor polynomial or some other Padé approximant for this function at a point will, when extrapolated, eventually behave very badly, due to the asymptotic behavior of polynomials. The well-behaved characteristics of the sine function come from scaling behavior, which necessarily extrapolates polynomial behavior, and comes from this extrapolation to an "infinite polynomial", i.e. the sine function.

The sine function is arrived at as a simple linear combination (over the complex field) of exponential behavior. While y = exp(x) displays the extrapolated growth characteristics of polynomials (growing faster than any polynomial), y = exp(ix) does not (where we assume x is restricted to real values). Any smooth (not necessarily analytic) extension of exp(ix) to the complex field displays growth or decay features.

We do not usually look at the exponential function in the following way, but it is an extrapolation of the polynomial

20 $\qquad p_n(x) = (1 + x/n)^n$

By selecting a branch of the nth root, we can define an inverse for this polynomial,

21 $\qquad T_n(x) = n\left(\sqrt[n]{x} - 1\right)$

Thus, while typical polynomial equations of arbitrary degree cannot be solved by general methods using radicals, an important point about the exponential-like polynomials is that whatever value we assign to $p_n(x)$, the resulting polynomial equation can be solved by radicals. Using different branches, we obtain all n solutions (possibly with multiplicity). We can look at $T_n$ as being a "multifunction", in that the n branches allow us to define locally n different smooth (or at least continuous) functions.

An elementary feature of (20) is that, for small x, while (20) will have a fairly high degree, as a polynomial, it behaves almost like a linear function. Thus, we have a function

22 $\qquad f(x) = A(x)$

And this function is almost linear for x close to 0. Also, we have a multifunction (21), which we write as

23 $\qquad g(x) = B(x)$

This function can be used, locally, to invert f:

24 $\qquad g(f(x)) = x$

(at least locally, selecting the appropriate "branches" for g).

Suppose A(x) is linear provided x is sufficiently close to zero. Then for x close to zero and λ real, A(λx) = λA(x), but this need not be the case if λx and x are not close to zero. What is the significance of this expression? If we know A(x) accurately, we can obtain an accurate estimate of A(λx), or vice versa (provided λ is not equal to zero). By (24),

25 $\qquad B(A(\lambda x)) = \lambda x$

Let us suppose that there is a (multifunction) S such that

26      $B(A(\lambda x)) = B(\lambda S(x))$

Then, this is very similar to

27      $A(\lambda x) = \lambda S(x)$

But this expression holds only locally, because B is just a local inverse.  However, when x gets close to 0, we will be able to approximate S as a perturbation on A,

28      $S \approx A + \Delta A$

This approximation, if used in (27), generally, will not be valid, because x need not be close to 0, where A becomes linear.

The point here is that although (28) is an approximation, it is based entirely on local information (near a point where there is an existing Taylor polynomial, namely x = 0, in this case.)  With this, (27) becomes:

29      $A(\lambda x) = \lambda (A + \Delta A)(x)$

This is only an approximation.  We can iterate this approximation,

30      $A(\lambda^2 x) = \lambda (A + \Delta A)(\lambda x)$
31      $A(\lambda^2 x) - \lambda (\Delta A)(\lambda x) = \lambda A(\lambda x) = \lambda^2 (A + \Delta A)(x)$
32      $A(\lambda^2 x) - \lambda (\Delta A)(\lambda x) - \lambda^2 (\Delta A)(x) = \lambda^2 A(x)$

But, the operators, once their arguments are sufficiently close to zero, have accurate approximations based on knowledge of the Taylor polynomial.  We will have to use the known form of ΔA to evaluate ΔA(x), and ΔA is only accurately known close to zero.  This means we are basing an evaluation of A(x) on an approximation obtained near 0 rather than (possibly) x.  We choose λ so that λx is shifted toward the optimal range near 0 where the Taylor polynomial can accurately assess values of A and ΔA.

By using the known expression for ΔA, we can, using iterations on (29), extrapolate the values of A beyond 0, based on the behavior of the Taylor polynomial near 0.  Near x = 0, we can, because the Taylor polynomial accurately approximates A, expand A in a Fourier decomposition.  This allows us to approximate ΔA in terms of Fourier components.  Because λ in (30), for example, will distribute over terms of ΔA, if we know how to carry out the type of transformation we have been discussing for trigonometric sines and cosines we will be able to use the type of extrapolation procedure we have been discussing.  Thus, the sine function is important to study under this type of scaling transformation.

If the extrapolation proves to be feasible, then we will have essentially converted the Taylor polynomial to a Fourier series.  This will provide a much more stable type of extrapolation than polynomials, due to the periodic nature of trigonometric functions.  Of course, there is the serious question of precisely what interval to assess the Fourier decomposition.  In addition, the extrapolations from each point at which a

Taylor polynomial is known will overlap, and also overlap with regions where the function is behaving nonlinearly. The question then arises as to what to do about the boundary regions. This is essentially problem 3, which we take up after we consider this problem more fully.

Let's consider Taylor polynomials generated at x = 0 for convenience, as we have in our discussion above. For the sine function, due to the identity sin(2x) = 2 sin(x) cos(x), we can do quite a bit better than the rough approximation we have described above. We have sin(x) = 4 sin(x/2) (d/dx)sin(x/2). This is an interesting identity, because it is telling us that if we know sin(x/2) very accurately in a neighborhood of x, that we can approximate sin(x) accurately, not using the behavior of the approximation near x, but near x/2. This is exactly the type of goal we had above, except instead of a rough approximation, we are getting an identity. Thus, let R be the approximation to sin(x). We can define a new approximation, $R_{new}$ by, $R_{new}$(x) = 4R(x/2) (d/dx)R(x/2). This modifies R in the vicinity of x, on the basis of the approximation in the vicinity of x/2, which might be much better-behaved. We say we are renormalizing R. We can define what we refer to as a renormalization transformation T by TR(x) = 4R(x/2)(d/dx)R(x/2). The previous approximation we discussed was much, much rougher than this renormalization transformation.

Let us imagine that sin(x) is not just the well-behaved trigonometric function we are accustomed to, but near x = 0 is altered, and the transformation is approximate rather than exact. The idea behind the renormalization transformation is that R might be a much better approximation to sin(x) near x/2 than near x, for a particular value of x. Then, on the large scale, we are merely utilizing an approximate extrapolation to the periodic function sin(x), whereas, on the small scale, we can get information about the true behavior of the function. Note that for the renormalization transformation, we are unconcerned about linearity. Although we are using both R and dR/dx, and therefore require significantly more information to know dR/dx accurately, that information might be available near x/2.

If it is the case that R is not a good approximation near x/2, we can apply T again, and consider $T^2R(x)$, which depends on x/4. Furthermore, if we do this to obtain a good approximation at x, we will have to have accurate approximations for R, dR/dx and $d^2R/dx^2$ at x/4, which, incidentally, we may possess, due to knowing R and some of its derivatives accurately at x = 0 (because we know a Taylor polynomial there). Clearly, provided we have the accurate information, we can just continue applying T. It does not take many iterations (relatively, because this is an exponential process) before $T^nR(x)$ brings $x/2^n$ very close to zero, and possibly into the region where we know R (and some of its derivatives) very accurately from the Taylor polynomial for sine. We need to know derivatives of R accurately up to $d^nR/dx^n$ at $x/2^n$.

There are a number of points to be made here. First, we may well know accurate derivatives for the function, here a function similar to sine, to very high order at x = 0. Second, there may be a fairly large microscopic region (but perhaps small macroscopic region) near x = 0 where machine limitations have not entered the picture, and where we are able to utilize this information from sine and its derivatives. A third point is that we derived the basic renormalization transformation from knowledge of sine. For a more empirically-derived function or for more complicated function, such a nice transformation may not be obvious at all, and our prior discussion remains germane. However, we are trying to extrapolate outward to a periodic sinusoid. All sinusoids possess some transformation property as we have discussed. If we utilize information accurately up to the nth derivative from a Taylor polynomial, the problem then resolves into identifying the sinusoid, or the range of sinusoids, that best fit the information from the Taylor polynomial. A fourth point is that the need for accurate knowledge up to $d^nR/dx^n$ is somewhat deceptive. This derivative is embedded as a factor in a single term, and therefore its importance is relative. It may be the case that due to the overall structure of the transformed expression, the highest derivatives do not contribute much. A last point is that, while T is unambiguously defined, $T^{-1}$ may not be. In other words, there may be a number of sinusoids that fit the data adequately. In this case, we must think of T as resulting from a product. Furthermore, R(x/2) and (dR/dx)(x/2) are entangled in this product. To define an inverse, we must disentangle these to yield R(x). This inverse problem may not be unambiguously solvable (despite the fact that R and dR/dx are not exactly independent). The transformation T is called more commonly a renormalization group transformation. But since these types of transformations are not necessarily invertible (and in fact, are likely to be non-invertible) the use of the term "group" is, as is well-known, somewhat of a misnomer. A question we have about this transformation, with respect to sine, is whether or not the 2 here is somehow characteristic, like an eigenvalue. In our previous discussion, using λ, there seemed to be no preferred value.

The renormalization group transformation (at least for sine), although nonlinear, is actually reminiscent of an eigenvalue equation for linear operators:

33
$$TR(x) = [2(d/dx)R(x/2)] \, 2R(x/2)$$

For a linear operator: 2R(x/2) would just equal R(x) and [2(d/dx)R(x/2)] is just the "slope" of the linear operator (since we are working with a single real variable x). The slope, rather than 2, is the "eigenvalue". Still, one wonders if 2 is somehow uniquely important.

This type of transformation acts on a function space, where R is transformed to TR. Therefore, where the scale change, x to x/2, must surely be emphasized, we see that the renormalization group transformation, when iterated, yields a discrete trajectory in this function space.

Let's point out something significant about (33). We are thinking in terms of R being a polynomial or a rational function. Thus, first, we note that our space of functions consists of rational functions, on which T acts. Secondly, if R is a polynomial of degree N, then TR is going to yield a higher degree polynomial (usually) and in fact, we can expect a polynomial of degree 2N-1. This is extremely significant, because as the degree of polynomials increase, we must expect them to become much more sensitive to slight changes in x. By focusing on x close to zero, we do not need to concern ourselves much with this, but this sensitivity is important, because the degrees of the polynomials grow exponentially as we iterate.

Let's examine other trigonometric functions besides sine, now that we see how ideal sine is for a change of scale transformation. We first consider a change in phase. Changing the phase of sine does not really add much in the way of challenge, until we obtain the cosine, because it varies quadratically, instead of linearly, at x = 0, and therefore is of a different nature. We still have, however, a nice identity that cos(x) is equal to $2\cos^2(x/2) - 1$. We are interested in following the line of development for sine. Therefore, we want the parabolic equivalent to the result for sine. First, we need to establish the appropriate initial condition, which is that the function equal zero.  Therefore, it seems better to consider $1 - \cos(x)$, which is equal to $2 \sin^2(x/2)$.  Thus, for R approximating $1 - \cos(x)$, we can define TR(x) to equal $2 (dR/dx)^2(x/2)$.  This, too, is a renormalization transformation, from a degree N polynomial, for example, to a degree $2N - 2$ polynomial.

The next case we must consider (with the idea of Fourier analysis in mind) is sin(kx), where k is some positive number.  But sin(kx) is just $2\sin(k(x/2))\cos(k(x/2))$.  Thus, we can readily define a renormalization transformation for this, also.

Next, we need to study the transformation of trigonometric series, as this is, fundamentally, what we are interested in.  Let's first look at a superposition of two sine functions.  The example sin(x) + sin(2x) is adequate.  This is equal to $2 \sin(3x/2) \cos(x/2)$, so we could define a transformation with respect to this.  But we have sin(x) + sin(2x) equal to $2 \sin(x/2)\cos(x/2) + 2 \sin(x) \cos(x)$, showing that a renormalization transformation is perfectly feasible, and can act linearly.

Now that we have seen how ideally-suited sines and cosines are for renormalization transformations, we can supply a solution to the second problem.  We do not simply want to expand the Taylor polynomial in a Fourier series, near x = 0, and then use periodicity.  While the trigonometric functions have ideal scaling properties, the function represented by the Taylor polynomial may not.  We must use the Taylor polynomial, in its domain about 0 where it is an accurate approximation to the function, to estimate ΔA of Eqn. (28), and use the methods there discussed to scale.

If ΔA is estimated from the behavior of the Taylor polynomial in an interval (a,b) about 0 where the polynomial is an accurate approximation, then for x outside this interval we use some value λ so that after n iterations $\lambda^n x$ is in this interval.  For example, if n = 2, we would use (32) to estimate A(x).  But: What value of λ should we use?  For the trigonometric functions, λ = 1/2 is satisfactory.  In (32), the end term, $\lambda^2 (\Delta A)(x)$ is presumably going to be one of the most inaccurate corrections.  This would seem to suggest that smaller values of λ are preferred to reduce the relative importance of error-generating terms such as this.  However, the smaller the value of λ, the faster $\lambda^n x$ enters the approximation region near 0 as n increases.  This defines a certain length scale for correlation between regions, and the actual correlation may be shorter than this, so that larger values of λ may be favored.

These extrapolated values avoid problems arising from nearby sharp edges, which, in using Fourier decomposition, result in the artifact of Gibbs phenomenon.  However, by using extrapolated regions, we do not "see" sharp edges at all.  Furthermore, we obtain extrapolation regions for each point at which we have a Taylor polynomial.  These regions will eventually overlap.  Thus, boundaries, sharp edges, and overlap regions become essential to consider.  It is interesting that this extrapolation approach is not subject to Gibbs phenomenon or the instability of polynomial approximation.  However, it certainly leaves us with some difficult considerations, addressed in our discussion of the third problem, below.

**Problem 3**

This problem addresses the aspects of data analysis for the mesoscopic wavefunction that Problems 1 & 2 have left open and unexplored.  Problem 1 was a problem directed toward unification of information from all Taylor polynomials at each point. We encountered problems with instability (a general problem associated with polynomials and rational funcions) that can be solved, in part, using non-analytic functions to damp instabilities. The resulting functions are akin to "standing waves", derived from information from the Taylor polynomials available at various points. Each Taylor polynomial is locally well-behaved, because much of the "noise" is present locally only for high derivatives. Taylor polynomials are probably not very sensitive to noise, and to local boundary behavior (that does not represent noise). However, beyond a local region of "adequate" approximation, the polynomials will display a great deal of instability, as they are attracted to the fixed point at infinity. Indeed, little about the unified polynomial, which is likely to have very high degree relative to the Taylor polynomials, or otherwise unified function can be expected to be stable, even when stabilized with non-analytic functions. This suggests the idea of introducing local "fixed" points for the polynomials. The reciprocal of the polynomials will display singularities at the zeroes of the polynomials. We can think of these singularities as fixed points, because change is largely orthogonal to the x-axis, and so we can think of positions as relatively stable. This will also be the case for poles of rational functions. Thus, with respect to changes in the x-direction, singularities stabilize the position and the low derivatives, projected onto that axis. Nevertheless, the overall significance of resolving problem 1 left us with a wavefunction that with respect to representation of uncertainty, was not wholly satisfactory. Unification of the information from the Taylor polynomials is helpful locally about the generating points of the polynomials, but otherwise displays too much instability, even when stabilized using non-analytic functions.

Problem 2 made some progress in terms of improving representation of uncertainty, although it takes an opposing perspective, building approximations by scaling outward from the local regions where the Taylor polynomials are good approximations. It is straightforward to outwardly scale trigonometric functions, because they display correlations on all scales, and because there are simple identities that permit remote regions to be connected to Taylor polynomials generated at any point. This ease is not readily apparent for Taylor polynomials in general. However, it does demonstrate that for some functions there is a direct connection between delocalized representations and the localized representation of Taylor polynomials. Extending the range of good approximation of a Taylor polynomial, in general, depends on developing local scaling relations, akin to perturbation expansions. This is a satisfactory approach, but the method we presented is a bit complicated. Developing scaling relations when correlations are not long range is complicated. Furthermore, the structural capabilities of such an approach are limited. One cannot predict intricate local structure, nor the position of boundaries, nor the precise structures involved in overlap between two extrapolations from Taylor polynomials generated at different points.

Clearly, Problems 1 & 2 leave much unanswered. Both problems try to solve the "inverse problem" for this situation: Take the information from the Taylor polynomials, and predict the structure globally of both information and uncertainty. This is a difficult complex task. The Taylor polynomials lack information about higher order derivatives. We must largely view the information used to construct the Taylor polynomials as phenomenological or empirical. This is far-removed from much of the microscopic theory of the wavefunction coming from quantum mechanics. Phenomenological information will not be just precisely what we want: There will be a conflation of a number of different types of information. Furthermore, data collection, however it is done, is not just going to represent accurately the signals we are interested in, but will contain some noise. A function, the existence of which is postulated theoretically, may not, strictly speaking, exist in a conventional sense, but perhaps only statistically. Furthermore, while such an object may exist, it may not be a function at all: We need to bear in mind that the concept "function" is a mere artificial construct. A further problem, not addressed by Problems 1 & 2, are machine limitations, such as the involvement of round-off errors and other types of problems. Thus, the approach to a direct attack on the "inverse problem" is intertwined with numerous difficulties, and it would be easy to cite more. For example, the discrete locations at which Taylor polynomials are defined does not supply correlated information necessarily with points remote from the generation points of the Taylor polynomials.

The inverse problem, itself, even if one overcomes the difficulties addressed in the previous paragraph, is an unwieldy problem. The point here is that presumably the various points of interest and the points at which Taylor polynomials are generated are interconnected. Furthermore, the interconnections may depend on other points. This means that potentially the inverse problem is an enormously complex network problem. Also, while the relationships may be relatively simple between nearby points, and even possibly linear, in general, we cannot expect anything so simple. With data arrived at empirically, as well, there is, at the very least, the possibility of important extraneous features, such as time variation, not being accounted for in our theoretical framework. The mesoscopic domain takes us within the realm of quantum mechanics, but it is within the classical limit of that theory. Therefore, as we decrease the number of points where we have classical information, i.e. the points at which we are generating classical information to construct the Taylor polynomials, Planck's constant must intrude at some point, but it is not clear how.

Rather than solving the inverse problem, we consider, in this third problem, the possibility of circumventing the inverse problem. Essentially, we are addressing a problem on a complete graph. The points of the graph are the positions of the Taylor polynomial generators, and the edges are the physical connections between Taylor polynomials. The function we are interested in is characterized at each point of the graph by a physical state, i.e. its Taylor polynomial to a certain degree, d, on that vertex, its generating position. We may have no information about the state at some points, and the mere location of some vertices could be uncertain (merely characterized by some defining criteria which indicates the importance of a Taylor polynomial generated from that point, but none has actually been generated.) However, there will be a certain set of vertices on which the states are known through the Taylor polynomials. At these points, we possess some physical information. On the other hand, as we have seen in our discussions of Problems 1 & 2, this information, in general, will be of very limited use in predicting the physical states at the other vertices.

We can think about training a neural net to circumvent the inverse problem. This is a sound alternative, but it is not likely to show us a path from our classical states to quantum mechanics. There has to be something limiting about the situation we have described, or something standard, in order to consider this alternative. In addition, by circumventing the inverse problem, we must expect, to a certain extent, statistics to play a role. Perhaps Planck's constant enters into the mesoscopic wavefunction characterization through this back door.

Consider the example of a unit step function. For x < a, Taylor polynomials generated from accurate physical information for the function have value 0 and derivative values of 0, while for x > a, the Taylor polynomials have value 1 and derivative values of 0.  In this simple type of problem, involving a boundary, we can estimate the boundary by using randomly selected points.  In fact, a uniform distribution would do satisfactorily.  There are a few difficulties, like merely finding an interval in which there is a step.

This type of consideration can get us started at circumventing the inverse problem.  One could pursue a few other elementary examples, but the statistical approach to circumventing the problem is relatively straightforward, and very likely to prove ultimately compelling in giving us a connection with quantum mechanics.  We have discussed the quantum mechanical wavefunctions in the introduction, and it is imperative that a mesoscopic wavefunction connect to this microscopic regime.  We are approaching this whole issue from construction of a wavefunction from classical information.  As we decrease the amount of classical (point) information, the regions of uncertainty become more significant, but they do not automatically lead us to an encounter with the microscopic regime and Planck's constant.  Problem 3 is clearly the most significant of the three problems, in attempting a rudimentary connection.  In other words, we are claiming that there is a unification that occurs, within the mesoscopic domain, between classical and quantum behavior.  A mesoscopic wavefunction, bearing classical information as well as information about uncertainty, is an interesting theoretical approach to unification, even if it does not have the framework, say, of a Schrödinger equation, and is not fundamental, but strictly empirical and phenomenological.

Briefly, our approach to this would be to treat the system of Taylor polynomials as "spin-like" structures for an Ising-like model.  Then we can define a classical Ising-like Hamiltonian with these "spins".  This artificial Hamiltonian is similar in structure to the Hubbard model, but strictly of a classical nature.  We will not pursue this idea further here, but this may provide a statistical inroad to a connection with quantum mechanics.  We leave consideration of the third problem now, in this inchoate state.

**Discussion**

The third problem was, of course, was the most important problem, and we progressed *very* little along that path, although there is an intriguing possibility that a mesoscopic wavefunction might open up a new perspective on the "classical limit".  Furthermore, our solutions to the first and second problems certainly pose various difficulties.  The solution to the first problem, while very straightforward, has a potential (which in our opinion would often be realized in practice) of resulting in a standing wave solution that displays a lot of instability.  The solution to the second problem, based as it was on relatively low degree Taylor polynomials (and of necessity so) misses important details such as boundaries, and has no satisfactory way of coping with overlap.  In the third problem, we suggest that a phenomenological or empirical approach, using a phenomenological Hamiltonian such as an Ising-like mode, based on learning and other phenomenological strategies, such as genetic programming, has a potential for circumventing the difficulties encountered in the first and second problems.  Because in these problems, our goal is to directly solve a large network problem, on the basis of some knowledge about states at particular vertices (which provide only very information, as this knowledge can be the result of conflation with macroscopic noise and other factors), we refer to the global problem that is being addressed as an inverse problem.  By focusing on learning or evolutionary behavior, we are suggesting in the third problem that it may be possible to satisfactorily circumvent this inverse problem.

However, such circumvention, in the abstract, involves a high degree of complexity. This goes beyond the goals of this paper.

It is important to critique the central and conventional idea of function. It would certainly be ideal, to have just a given function to consider, say from a model, or from experiment or observation. However, a given, well-defined function may well be only the end product of the sampling process involved in a model or experiment. Because of the discrete nature of a sampling process, and because this often amounts to a time series, our function (regarded as empirically defined) F, of interest for approximation and representing information, is better-described as a function of discrete time n = 0, 1, 2, …, and there is now a need to associate with a given n, the position x(n) of interest. In this way, strictly, as a "function" of x, we can define a random variable f(x), which at a given x, has associated pairs (x(n), F(n)), where x(n) = x. As a random variable, f(x) then takes the value F(n) with a certain probability. With this definition, we can restrict x to possible values of x(n), and regard it as a discrete variable, defined with respect to finite lattices in one-dimension, for example. With such an indirectly-defined function, what becomes of the notion of Taylor polynomial?

Our suggestion for this is to have associated with x, two graphical structures: A macroscopic, but small lattice, and a mesoscopic (meaning bordering on the microscopic and perhaps extending into the microscopic) graph, with a "cloud" of x-values connected to each value of the macroscopic lattice, and representing points about a particular lattice point X that can be used to define a Taylor polynomial at X for the random variable f. Then, the mesoscopic graph on x-values is a random graph, while the macroscopic lattice of points is not. The values x(n) are drawn from the macroscopic lattice, while the actual sampled point X(n,m) is a random variable on a time m associated with the mesoscopic graph. When the macroscopic time is a particular value of n, the time m = 0, 1, 2, …. M, changes to some maximum value M, and represents the actual sampling of x at the lattice value x(n). The sampled values F(n) constitute the measured values for the function at these m values. Since we need to construct a "Taylor polynomial" from these measured values, we need a random variable of values associated with the "scatter" X(n,m) about x(n). This means that while we are able to define the random variable f(x) for x = x(n), there must be another scale, the mesoscopic scale, from which we draw values for a "Taylor polynomial".

We expect that, due to randomness, the degree of a Taylor polynomial that one is able to define for a random variable will be very constrained. However, as we have mentioned previously, it is not necessary to push the degrees to high orders, as the level of noise in data becomes more significant with increased information needed to estimate higher derivatives. A capability of computing estimates of position (some type of local mean for f), velocity (a local "macroscopic" derivative) and acceleration (again a macroscopic quantity, representing a second derivative estimate from the random variable f), in fact, is a significant amount of information, especially if a large number of macroscopic lattice points are sampled (at a mesoscopic or local level).

**Gibbs Phenomenon**

A final point to address about our work on these three problems concerns Gibbs phenomenon. We have provided a start on the way of possibly circumventing Gibbs phenomenon by an empirical approach that is local, depends on scaling, and infers sharp boundaries, without entailing any artifact like Gibbs phenomenon as occurs in Fourier analysis.

Circumventing artifacts in noise might result in some significant improvements in applications in which persistent non-local artificial noise such as Gibbs phenomenon appears. Of course, there has been much progress in this area in research, and we are not claiming that we have a competitive approach, merely that we have made significant progress, especially on the Problems 1 & 2, shifting the viewpoint away from analytic functions and Fourier analysis to renormalization. Renormalization methods utilize local information, which has the potential of delineating boundaries. These methods extend and extrapolate via scaling, and it is the overlap between local extrapolated regions which might be addressable, especially in certain model systems.

The availability of information about Taylor polynomials at various points can be regarded as analogous to information about spin in the Ising model. By extrapolating, one has the potential of constructing an interaction model, through knowledge of overlap between two extrapolations from Taylor polynomials. This would serve to estimate interaction between the "spins", i.e. the Taylor polynomials. By converting the problem into an Ising-like model, one can then delineate domains, with boundaries at the overlap in domains fairly well defined. As this model is phenomenological and based entirely on data, as we indicate in the discussion section, it can be regarded as a dynamical model. Although it is unusual to focus on kinetic Ising models, the approach we are suggesting from addressing the three problems discussed in this paper, might provide a definite alternative dealing with Gibbs phenomenon. The boundary domains in Ising-like models can be worked out in simulation, and would result in circumventing Gibbs phenomenon.

**Conclusions**

We have discussed three problems related to a description of mesoscopic information. A function, f, of interest is considered, defined on a real domain, and considered to be a $C^\infty$ function on this domain. This approach is based on examining Taylor polynomials generated at various positions. The first two problems represent direct attacks on obtaining global information from the local information supplied by the Taylor polynomials. We view this information in the context of Ising-like models in statistical mechanics. Each Taylor polynomial, at the site of its generation, represents a spin-like structure. So, at a position x, f(x) is considered a "position", f'(x) a "velocity", f''(x) an "acceleration", etc. in an empirical context, and the spin-value at position x is the Taylor polynomial constructed from these mechanical values. One can envision a one-dimensional finite lattice with very many relatively low-degree Taylor polynomials at each site. It is then of interest to get a picture of the overall magnetization that these polynomials define. This is the role of the first two problems. We regard this magnetization as entirely structural and based on the spin-like Taylor polynomials at each site. Therefore, the overall problem being addressed in the first two problems is to obtain a picture of this magnetization, as a mesoscopic wavefunction. The first problem regards the Taylor polynomials as essentially yielding local information, as well as uncertainties in a global picture, and the second problem poses the question of obtaining overlap about these local islands of information that can be used to define coupling constants for spin-spin interactions of the Ising-like model of these Taylor polynomial spins. In discussing the first two problems, we ignore fluctuations, and reduce our considerations to a simple mesoscopic system, rather than focusing on microscopic fluctuations and correlations. It is, however, precisely this type of information that is likely to be available as measuring devices are miniaturized, so such a phenomenological approach, accepting a conflation and smoothing of microscopic fluctuations, is likely to be meaningful in applications.

The solution we obtained to the first problem is compromised by the instability of high-degree polynomials. This might be thought to be a difficulty that one does not need to face because, at each point of the lattice, only relatively low-degree mechanical information is available, and the instabilities yield justifiable regions of uncertainty for the mesoscopic wavefunction. However, in obtaining a unified picture, very high degree polynomials become important and one encounters rapid departures from accuracy when one moves beyond the local region about each point where the Taylor polynomials represent satisfactory approximations. We have discussed the stabilization of this by using smooth non-rational functions. This relates to renormalization, which is only discussed in the context of the second and third problems. In this paper, we have definitely made a start at developing a satisfactory theory in the mesoscopic domain.

While numerous infinite series display very stable, even periodic patterns, the finite truncations of such series in Taylor polynomials at high degree display significant instability, and are subject to polynomial wiggle with an attraction to infinity. The first problem leaves us within this context of a region of divergence in a mesoscopic wavefunction on the global scale. About each Taylor polynomial, we are aware of the islands of good approximation, surrounding the generating point of each Taylor polynomial. Furthermore, it is clear that simply focusing on these islands misses the possibility of extrapolations, which while not especially accurate may be expressive of scaling trends in the Taylor polynomials. The point is that each Taylor polynomial can be extrapolated to suggest an approach to macroscopic behavior.

Some functions that we can use to describe the ideal behavior of an ordered Ising mesoscopic state will not display instability. These ordered mesoscopic states can be decomposed approximately into sinusoidal-like functions. For functions that are periodic, like sine, these large scale trends are expressive of certain identities that characterize the functions. Order is locked in at every length scale. Utilizing these identities, we can extrapolate Taylor polynomials rather successfully beyond the range in which they are strictly well behaved by using the identities as renormalization transformations. The sine function can be taken to approximate a component of a mesoscopic wavefunction displaying long-range order with respect to correlations. Thus, we can regard its periodic structure abstractly, and part of a structure that, at very small scales, may diverge significantly from the scaling structure being described by a renormalization transformation.

Mesoscopic states characterize and summarize the results of making many measurements, gradually encroaching from the macroscopic to microscopic domains. The measurements are taken on a very small time scale, and locally, they are not purely macroscopic. The result is a mix of near-classical behavior, with expressions of high level of uncertainty between these sites. This is intermediate between quantum and classical behavior for small systems, and reflective of a different picture of the "classical limit" than conventional. The global picture of a mesoscopic wavefunction is reflective of the conflation of many macroscopic measurements, but not enough to present a purely statistical picture of a macroscopic state.

The role of the second problem is one of extending the local macroscopic measurements represented by Taylor polynomials. The Taylor polynomial at a point will not be like a sine function, but can be scaled by a renormalization transformation. The characteristic periodicity for sine is an expression of a component from this scaling procedure. Each Taylor polynomial has some region about which it supplies good approximations from classical, macroscopic measurement. Through renormalization transformations, we can capture an extended picture in a region of uncertainty.

We must extrapolate from the classical-like regions at the site of Taylor polynomials, lacking global certainty of a macroscopic state. This is the goal of the second problem. We have used the renormalization transformation to provide us with a method for extrapolation. Because there is uncertainty about the surroundings, the extrapolation is not meaningful either macroscopically or microscopically. It is reflective of ambiguities of scaling extrapolations. Vaious possible scaling extrapolations result, when put together, in a single mesoscopic state.

The extrapolations provide us with overlap regions. We do not develop the idea, but we can use the overlap to elaborate on a simple Ising-like model. We can estimate a coupling strength between two Taylor polynomial extrapolations, with the Taylor polynomials themselves regarded as spin analogues. In order to do this, we must characterize the structure of the spin, which is a "classical" quantity. A Taylor polynomial at a site constitutes a vector structure not a scalar spin, with components being its value at the site, the value of the derivative at the site, the value of its second derivative at the site, etc. We must compare like components A and B from Taylor polynomials generated at two nearby positions: for example as the size of $(A - B)^2$. Then $A^2$ and $B^2$ are always positive, so the only quantity that can express a relative difference between the two components is AB. We need to compare extrapolations of the Taylor polynomials. Since the principle is the same to whatever order we take derivatives, we can think of comparing Taylor polynomials just to first order at two sites x = a and x = b, namely f(a) with f(b)+f'(b)(a-b) and f(b) with f(a)+f'(a)(b-a). But, these simple extrapolations are not reflective of the scaling. Instead, we need to compare f(a) with $f_{ext}$(a;b,f(b),f'(b)) and f(b) with $f_{ext}$(b;a,f(a),f'(a)). Here $f_{ext}$(x;a,f(a),f'(a)) signifies the value of the extrapolation at x based on the spin configuration f(a), f'(a) at the lattice site x = a. Then the quantity AB we mention above is going to be taken to equal some combination of f(a)$f_{ext}$(a;b,f(b),f'(b)) and f(b)$f_{ext}$(b;a,f(a),f'(a)). The coupling strength used here must depend on the overall connection between the two sites arrived at from these quantities. We suggest using the resulting global approximation presented as the solution of Problem 1. We can integrate the square differences between the global wavefunction and each extrapolation between x = a and x = b as a measure of the reciprocal of the coupling, in each case of the combinations presented above. For example, for the coupling strength for the spin product f(a) $f_{ext}$(a:b,f(b),f'(b)), we would use the reciprocal resulting from comparing $f_{ext}$(x;b,f(b),f'(b)) to the global approximation. We can combine information locally in a linear way at each site x = a. This gives a local Ising-like spin-spin coupling Hamiltonian, necessarily localized to each Taylor polynomial generating site.

We are going to generate Ising-like couplings, necessarily, in this way. Near a boundary overlap between two Taylor polynomial extrapolations, couplings will reflect large local discrepancies in extrapolations. In a fully-elaborated Hamiltonian, statistical methods, such as Monte Carlo, would result in domains, and there is no Gibbs phenomenon.  On the other hand, the macroscopic lattice we use initially may be too crude to map domains very well and indicate boundaries.  One can then refine the lattice, retaining the old lattice sites (and therefore not disposing of the information from these sites) and add new sites to form a lattice with a smaller lattice constant.  This necessarily reduces the amount of microscopic information available.  This means that there will be limitations on how many refinements we can carry out.  But these limitations are empirical, and limited by the available data, machine limitations, and other environmental limitations, and are limitations one must expect.  Nor, at any point in a refinement, is there any need to confront Gibbs phenomenon.

In this paper, we have addressed three theoretical problems involved in developing a statistical mechanical approach to the construction of a mesoscopic wavefunction.  It is phenomenological and empirical and oriented entirely toward applications, rather than specific theoretical examples, although we have shown that it is related to renormalization.  This paper represents merely a first step in developing a method, and we have briefly outlined, in these concluding remarks, how further development might proceed.  The idea of a mesoscopic wavefunction is a complex one, and we have discussed it mathematically and abstractly in a simple setting of Taylor polynomials.  A mesoscopic wavefunction is describing a bridge between macroscopic and microscopic procedures, and for small systems, a transition from classical to quantum domains.  We have indicated that it would have applications beyond this to many other types of systems, such as DNA.